\documentclass{pasj01}

\begin{document}
\SetRunningHead{Toshiki Saito et al.}{Radio-to-FIR SED of NGC 1614 with ALMA and VLA}
\Received{2015/09/10}
\Accepted{2015/12/16}

\title{Spatially-resolved Radio-to-Far-infrared SED of the Luminous Merger Remnant NGC~1614 with ALMA and VLA}

%


 \author{%
   Toshiki \textsc{Saito},\altaffilmark{1,2}
   Daisuke \textsc{Iono},\altaffilmark{2,3}
   Cong \textsc{K. Xu},\altaffilmark{4}
   Junko \textsc{Ueda},\altaffilmark{2}
   Kouichiro \textsc{Nakanishi},\altaffilmark{2,3}
   Min \textsc{S. Yun},\altaffilmark{5}
   Hiroyuki \textsc{Kaneko},\altaffilmark{6}
   Takuji \textsc{Yamashita},\altaffilmark{7}
   Minju \textsc{Lee},\altaffilmark{1,2}
   Daniel \textsc{Espada},\altaffilmark{8}
   Kentaro \textsc{Motohara},\altaffilmark{9}
   and
   Ryohei \textsc{Kawabe}\altaffilmark{1,2,3}
   }
 \altaffiltext{1}{Department of Astronomy, The University of Tokyo, 7-3-1 Hongo, Bunkyo-ku, Tokyo 133-0033, Japan}
 \email{toshiki.saito@nao.ac.jp}
 \altaffiltext{2}{National Astronomical Observatory of Japan, 2-21-1 Osawa, Mitaka, Tokyo, 181-8588, Japan}
 \altaffiltext{3}{The Graduate University for Advanced Studies (SOKENDAI), 2-21-1 Osawa, Mitaka, Tokyo, 181-0015, Japan}
 \altaffiltext{4}{Infrared Processing and Analysis Center (IPAC), California Institute of Technology, 770 South Wilson Avenue, Pasadena, CA 91125, USA}
 \altaffiltext{5}{Department of Astronomy, University of Massachusetts, Amherst, MA 01003, USA}
 \altaffiltext{6}{Nobeyama Radio Observatory, Minamimaki, Minamisaku, Nagano 384-1305, Japan}
 \altaffiltext{7}{Institute of Space and Astronautical Science, Japan Aerospace Exploration Agency, 3-1-1 Yoshinodai, Sagamihara, Kanagawa 229-8510, Japan}
 \altaffiltext{8}{Joint ALMA Observatory, Alonso de C\'{o}rdova 3107, Vitacura, Casilla 19001, Santiago 19, Chile}
 \altaffiltext{9}{Institute of Astronomy, The University of Tokyo, 2-21-1 Osawa, Mitaka, Tokyo 181-0015, Japan}

\KeyWords{Galaxies: individual (NGC~1614, Arp~186, IRAS F04315-0840)  ---  Galaxies: starburst  ---  Radio continuum: galaxies  ---  Submillimeter: galaxies} 

\maketitle

\begin{abstract}
We present the results of Atacama Large Millimeter/Submillimeter Array (ALMA) 108, 233, 352, and 691~GHz continuum observations and Very Large Array (VLA) 4.81 and 8.36~GHz observations of the nearby luminous merger remnant NGC~1614.
By analyzing the beam (1\farcs0 $\times$ 1\farcs0) and $uv$ ($\geq$ 45~k$\lambda$) matched ALMA and VLA maps, we find that the deconvolved source size of lower frequency emission ($\leq$ 108~GHz) is more compact (420~pc $\times$ 380~pc) compared to the higher frequency emission ($\geq$ 233~GHz) (560~pc $\times$ 390~pc), suggesting different physical origins for the continuum emission.
Based on an SED model for a dusty starburst galaxy, it is found that the SED can be explained by three components, (1) non-thermal synchrotron emission (traced in the 4.81 and 8.36~GHz continuum), (2) thermal free-free emission (traced in the 108~GHz continuum), and (3) thermal dust emission (traced in the 352 and 691~GHz continuum).
We also present the spatially-resolved (sub-kpc scale) Kennicutt-Schmidt relation of NGC~1614.
The result suggests a systematically shorter molecular gas depletion time in NGC~1614 (average $\tau_{\rm{gas}}$ of 49 - 77~Myr and 70 - 226~Myr at the starburst ring and the outer region, respectively) than that of normal disk galaxies ($\sim$ 2~Gyr) and a mid-stage merger VV~114 (= 0.1 - 1~Gyr).
This implies that the star formation activities in U/LIRGs are efficiently enhanced as the merger stage proceeds, which is consistent with the results from high-resolution numerical merger simulations.
\end{abstract}

\section{INTRODUCTION}
Gravitational interaction between two gas-rich galaxies not only changes the morphology, but it also induces bursts of star formation (starburst; SB) (e.g., \cite{hop06, tey10}).
It is observationally known that most of local and high-redshift galaxies with elevated infrared luminosity (i.e., SB enshrouded by dust) are merging and interacting galaxies (e.g., \cite{kar10}).
Galaxies with total infrared luminosities ($L_{\rm{IR}}$) of $\geq$ 10$^{11}$ $L_{\odot}$ are called (ultra-)luminous infrared galaxies (U/LIRGs).
Since they are known to harbor nuclear activities (active galactic nucleus (AGN) and/or SB) hidden behind large column of dust, submillimeter wavelength observation is essential in order to investigate the obscured activities directly.

In the radio to far-infrared (FIR) regime, the spectral energy distribution (SED) can be modeled with three components, non-thermal synchrotron, thermal bremsstrahlung (free-free), and thermal dust continuum emission (e.g., \cite{cnd92}).
The synchrotron radiation in galaxies arises from relativistic electrons generated by shocks mostly associated with supernova remnants.
The free-free emission comes from H$_{\rm{II}}$ regions containing ionizing stars.
The flux of the free-free emission is difficult to measure directly, because the flat spectrum (index = $-$0.1 in the optically thin regime) makes it weaker than the synchrotron emission (index $\sim$ $-$0.8) below 30~GHz.
The FIR continuum is represented by a modified blackbody spectrum which is emitted by interstellar dust grains heated by stars to temperatures between 20 - 200~K.
High-resolution multi-wavelength observations are required to distinguish these three components.
Past studies of the radio-to-FIR SED for U/LIRGs (e.g., \cite{u12}) had focused on the global properties (e.g., total star formation rate (SFR)) due to the limited angular resolution at higher frequency ($>$ 100~GHz).
In this paper, we focus on the sub-kpc properties of the radio-to-FIR SED of a luminous merger remnant NGC~1614 using ALMA and VLA.

NGC~1614 is one of the nearby ($D_{\rm{L}}$ = 68.6~Mpc; 1\arcsec = 330~pc) gas-rich LIRGs (log ($L_{\rm{IR}}$/$L_{\odot}$) = 11.65; \cite{arm09}) with a tidal tail and an arc-like structure traced in the H$_{\rm{I}}$ \citep{h&y96}.
Numerical simulation by \citet{vai12} suggested that the galaxy is formed by a gas-rich minor merger with the mass ratio of $\sim$ 1/4.
Low-$J$ CO observations revealed that the system has a nuclear gas ring with an extended rotating gas disk and an arm-like structure \citep{ols10, kng13, slw14}.
NGC~1614 is classified as a merger-remnant with a cold gas disk \citep{ued14}.
The starburst ring is detected in the 5~GHz and 8.4~GHz radio continuum, Paschen $\alpha$, PAH, and CO~(6--5) emission \citep{aln01, ols10, hi14, xu15}, indicating a presence of strong SB along the ring (starburst ring).
The radio and X-ray studies revealed little evidence for the presence of AGN \citep{hi14}.
This is consistent with the results from an AGN diagnostics using the HCN~(4--3)/HCO$^+$~(4--3) line ratio \citep{i&n13}.
In addition, \citet{xu15} argued that the non-detection of the nucleus in the ALMA 435~$\mu$m continuum image ruled out any significant AGN activity.
We regard NGC~1614 as a starburst-dominated galaxy throughout this paper.

This paper is organized as follows: the observations and data reduction are summarized in Section 2, results are briefly summarized in Section 3, and modeling procedure of the SED is described in Section 4.
After discussing the spatially-resolved Kennicutt-Schmidt relation of NGC~1614 and other U/LIRGs in Section 5, we summarize and conclude this paper in Section 6.
We have adopted H$_0$ = 70 km s$^{-1}$ Mpc$^{-1}$, $\Omega_{\rm{m}}$ = 0.3, and $\Omega_{\rm{\Lambda}}$ = 0.7.

\section{OBSERVATION AND DATA REDUCTION}

\subsection{ALMA Observations}
Observations toward NGC~1614 were carried out as an ALMA cycle 2 program (ID = 2013.1.01172.S) using thirty-five 12~m antennas.
The band 3 and 6 receivers were tuned to the $^{12}$CO~(1--0) and $^{12}$CO~(2--1) line emission in the upper sideband.
The band~3 data were obtained on August 30, 2014 (on-source time of $T_{\rm{integ}}$ = 16.9 min.) using the projected baseline lengths of 28 - 1060~m.
The band~6 data were obtained on December 8, 2014 ($T_{\rm{integ}}$ = 22.0 min.) using the projected baseline lengths of 15 - 349~m.
Each spectral window had a bandwidth of 1.875 GHz with 3840 channels, and two spectral windows were set to each sideband to achieve a total frequency coverage of $\sim$ 7.5 GHz.
For the band~3 and 6 observations, J0423-0120 was used for the bandpass and phase calibrations, and Uranus was used for the flux calibration.
The main target lines of the ALMA program were the $^{12}$CO transitions (Saito et al. in preparation), and we subtracted all line features in the bandpass in order to obtain the continuum emission (Section~\ref{data}).

\begin{figure*}
 \begin{center}
	\includegraphics[width=16cm]{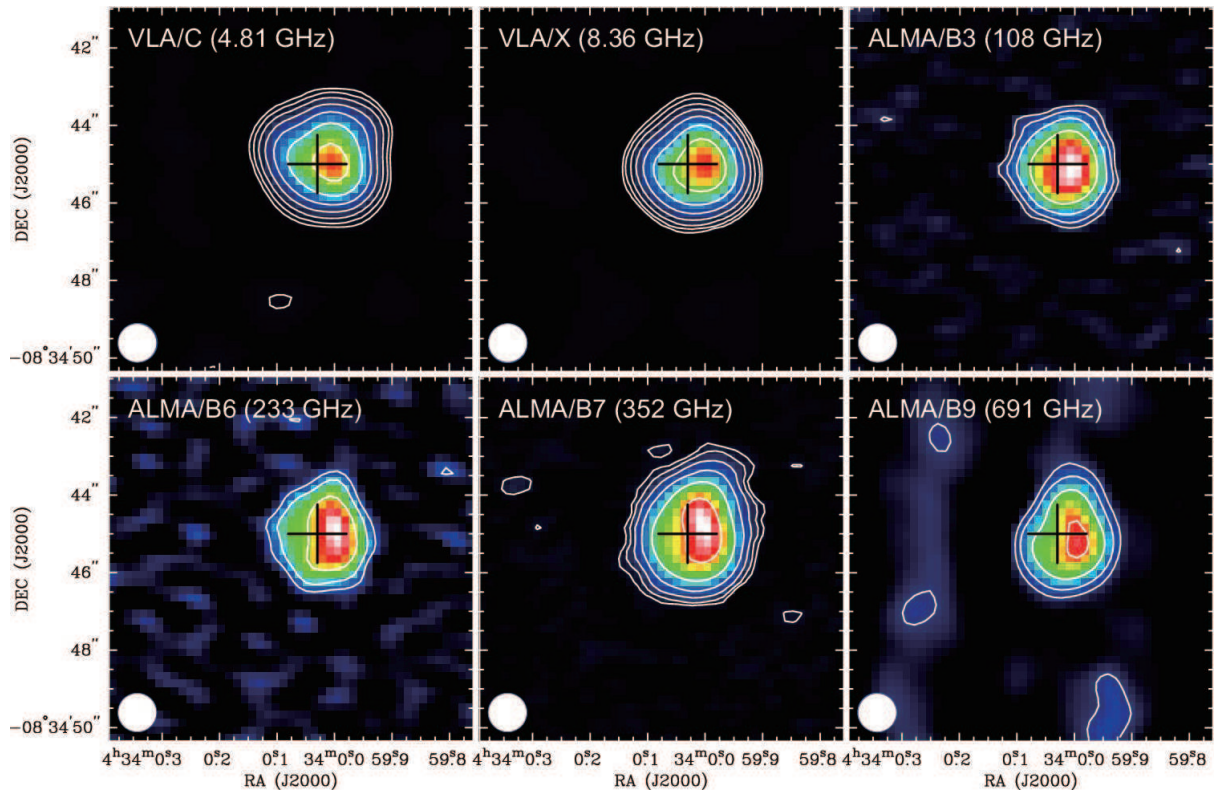}
 \end{center}
\caption{The 4.81 GHz to 691 GHz continuum images of NGC~1614.  The minimum $uv$ ranges of all data are clipped at 45 $k\lambda$ to have the same MRS of 4\farcs6.  The convolved beams (1\farcs0 $\times$ 1\farcs0) are shown in the lower left.  The $n$th contours are at 3$\cdot$2$^{n - 1}\sigma$ except for the 691~GHz image (2$^{n}\sigma$) ($n$ = 1, 2, 3 ...).  The rms noise are 45, 25, 51, 143, 68, and 4800 $\mu$Jy beam$^{-1}$ for VLA/C, VLA/X, ALMA/B3, ALMA/B6, ALMA/B7, and ALMA/B9, respectively.  The black cross indicates the peak position (nucleus) which is provided by the high-resolution image of the radio continuum \citep{ols10}.
}\label{fig_1}
\end{figure*}

\subsection{Archival ALMA and VLA Data}
NGC~1614 was observed using ALMA band~7 and band~9 as two cycle~0 programs (ID = 2011.0.00182.S and 2011.0.00768.S).
We obtained the calibrated archival visibility data from the ALMA archive (see \cite{slw14} and \cite{xu15} for the details of the data).
The band~7 data were obtained on July 31 and August 14, 2012 ($T_{\rm{integ}}$= 88.8 min.) with a projected baseline length of 18 - 1341~m.
The band~9 data were obtained on August 13 and 28, 2012 ($T_{\rm{integ}}$ = 24.7 min.) with a projected baseline length of 20 - 394~m.
J0423-0120 was used for the bandpass and phase calibrations for the band~9 observations.
J0522-364 was used for the bandpass calibration, while J0423-013 was used for the phase calibration for the band~7 observations.
For the band~7 and 9 observations, Callisto and Ceres were used for the flux calibration, respectively.

NGC~1614 was observed using VLA C-band (4.89 GHz) and X-band (8.49 GHz).
We use the archival calibrated visibility data (see \cite{cnd90}, \cite{sch06}, \cite{ols10}, and \cite{hi14} for the details of the data).
The C-band observations were carried out using twenty-four to twenty-seven 25~m antennas.
The C-band data were obtained on May 15, 1986 (A config.), August 10, 1986 (B config.), November 23, 1987 (B config.), June 17, 1994 (B config.), September 16, 1998 (B config.), and July 27, 1999 (A config.).
The combined data has the projected baseline lengths of 137 - 36,605~m.
The X-band observations were carried out using twenty-five to twenty-seven 25~m antennas.
The X-band data were obtained on August 3, 1993 (C config.), July 27, 1999 (A config.), July 9, 2001 (C config.), July 8, 2003 (A config.), October 15, 2003 (B/A config.), and November 6, 2004 (A config.).
The combined data has the projected baseline lengths of 32 - 36,620~m.
For the C-band observations, 3C48, 3C138, J0134+329, and J0420-013 were used for the flux calibration, and J0019-000 and J0423-013 were used for the phase calibration.
For the X-band observations, 3C48 and J0134+329 were used for the flux calibration, and J0423-013 was used for the phase calibration.

\begin{figure*}
 \begin{center}
	\includegraphics[width=16cm]{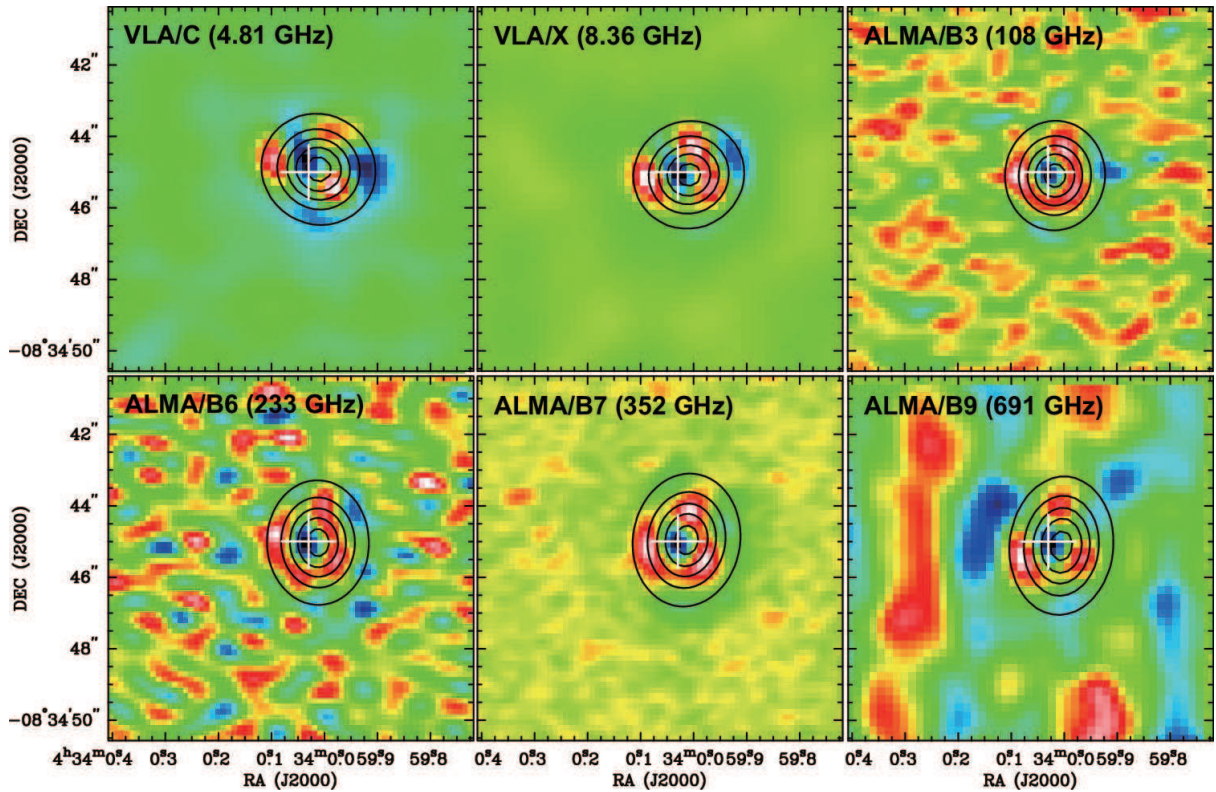}
 \end{center}
\caption{Model (contour) images and residual (color) images of the 2D gaussian fitting.  The contours are 10, 30, 50, 70, and 90\% of the peak flux.  The peak values are 14.3, 9.4, 2.6, 2.9, 9.1, and 95.6 mJy beam$^{-1}$ for VLA/C, VLA/X, ALMA/B3, ALMA/B6, ALMA/B7, and ALMA/B9, respectively.  The residual in color scale ranges from the minimum to maximum pixel value.  The minimum values are $-$1.0, $-$1.0, $-$0.4, $-$0.6, $-$1.9, and $-$17.4 mJy beam$^{-1}$ while the maximum values are 1.2, 0.7, 0.2, 0.5, 1.1, and 15.1 mJy beam$^{-1}$ for VLA/C, VLA/X, ALMA/B3, ALMA/B6, ALMA/B7, and ALMA/B9, respectively.  The white cross indicates the peak position (nucleus) which is provided by the high-resolution image of the radio continuum \citep{ols10}.
}\label{fig_2}
\end{figure*}

\subsection{Data Reduction and Imaging} \label{data}
For analyses in this paper, we assume that the missing flux effect (i.e., absence of short spacings) is negligible because we only discuss the structure that is smaller than the ``maximum recoverable scale" (MRS) of each observation.
This is given by,
\begin{equation}
{\rm MRS} \approx 0.6 \times \frac{\lambda_{\rm{obs}}}{L_{min}}
\end{equation}
where $\lambda_{\rm{obs}}$ is the observed wavelength and $L_{\rm{min}}$ is the minimum baseline in the array configuration\footnote{https://almascience.nao.ac.jp/documents-and-tools/cycle-2/alma-technical-handbook}.
The MRS is estimated from the minimum baseline length of the assigned configuration and the observed frequency.
Since the minimum baseline of all data shown in this paper is clipped at 45~k$\lambda$ (i.e., minimum baseline of the band~9 observation) before imaging processes, the produced images have a same MRS of 4\farcs6 ($\simeq$ 1.5~kpc).
The image properties (e.g., MRS, $uv$ weighting, and beam size) are listed in Table \ref{table_contin}.

We used the delivered calibrated $uv$ data for the band 3 and 6 observations, and the calibrated archival $uv$ data for the others.
The data processing was accomplished using {\tt CASA} (version 4.2.2; \cite{CASA}).
We convolved all images into a same beam (1\farcs0 $\times$ 1\farcs0).
We subtracted all strong line emission features (e.g., CO and CN) before the imaging of continuum emission by masking the velocity range of $\pm$ 500~km s$^{-1}$ centered on their rest frequencies.
This velocity range is consistent with the velocity of a putative CO~(1--0) outflow ($|v - v_{\rm{sys}}|$ $<$ 420~km s$^{-1}$; \cite{gar15}).
The rms level of 4.81, 8.36, 108, 233, 352, and 691~GHz images (Figure~\ref{fig_1}) are 32, 7.4, 51, 143, 68, and 736 $\mu$Jy, respectively.
The systematic errors of absolute flux calibration are estimated to be 3\%, 3\%, 5\%, 10\%, 10\%, and 15\% for 4.81, 8.36, 108, 233, 352, and 691~GHz, respectively.
Throughout this paper, the pixel scales of all images are set to 0\farcs2/pixel.
The absolute positional accuracy of the ALMA and VLA images are estimated to be less than 10\% of the synthesized beam size.
Since the synthesized beam sizes before convolving into the 1\farcs0 $\times$ 1\farcs0 resolution are smaller than 1\farcs0, the positional accuracy is better than 0\farcs1 which is smaller than the pixel size.

\begin{figure*}
 \begin{center}
	\includegraphics[width=16cm]{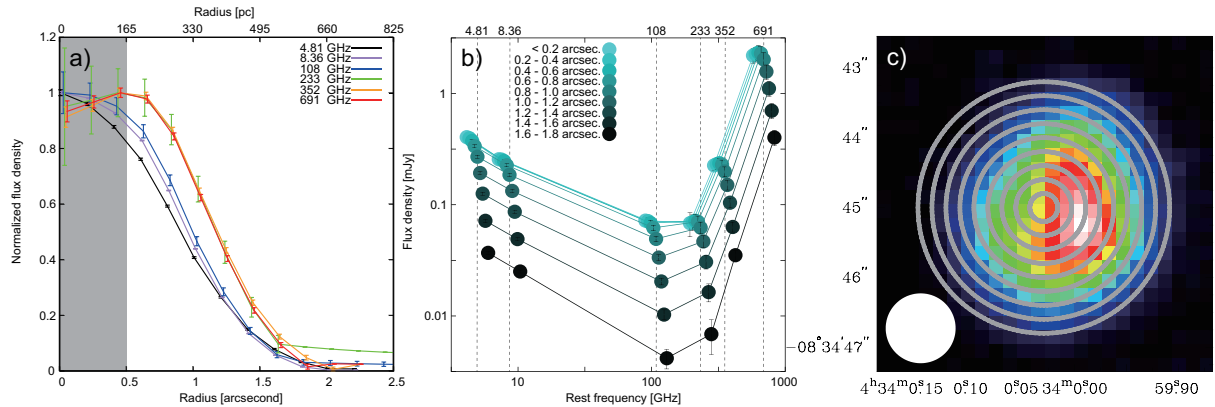}
 \end{center}
\caption{(a) Radial distribution of each emission. The shaded area indicates the convolved beam size.  We only consider the statistical error in this plot because the y-axis is the normalized flux density. (b) Spatially-resolved SED.  The color scale ranged from 0\farcs0 to 1\farcs8 indicates the radius of concentric rings which are shown in the right figure.  (c) Concentric rings which are used for the photometry overlaid on the 691~GHz image.  The grey solid rings indicate photometric regions where all the six continuum emission are detected. The convolved beam is shown in the bottom left corner.
}\label{fig_3}
\end{figure*}

\section{RESULTS}
\subsection{4.81, 8.36, and 108~GHz Emission}
The continuum images at 4.81, 8.36, and 108~GHz show a nearly circular (face-on) morphology (Figures \ref{fig_1}a, \ref{fig_1}b, and \ref{fig_1}c) with a deconvolved major axis and minor axis of 1\farcs28 $\pm$ 0\farcs03 ($\sim$ 420~pc) and 1\farcs15 $\pm$ 0\farcs05 ($\sim$ 380~pc), respectively.
The source size is derived by fitting a 2D gaussian in the image plane using {\tt imfit} in {\tt CASA} as shown in Figure \ref{fig_2}.
The fitting parameters are listed in Table \ref{table_gauss}.
The strongest peaks coincide with each other, and the peaks are $\sim$ 0\farcs8 west of the nucleus identified from the high resolution 5~GHz continuum using Multi-Element Radio Linked Interferometer Network (MERLIN) (synthesized beam of 0\farcs2; \cite{ols10}) and Paschen $\alpha$ image using HST/NICMOS \citep{aln01}.
The offset arises due to the asymmetric starburst ring structure \citep{hi14, xu15}.
All of the residual images show a same strong negative peak at the nuclear position and a same positive ring-like structure, suggesting all continuum images contain a similar structure.
This structure may be related to the presence of the starburst ring which is seen in all high-resolution multi-wavelength images \citep{aln01, ols10, kng13, hi14, xu15}.

A comparison of the 4.81~GHz flux observed by the VLA with the 4.83~GHz flux obtained by the Green Bank 91~m (GBT) telescope \citep{bic95} suggests that the VLA data recovers 42 $\pm$ 8 \% of the total emission (Table \ref{table_contin}).

\subsection{233, 352, and 691~GHz Emission}
The continuum images at 233, 352, and 691~GHz show a slightly elongated elliptical morphology from the south to the north (Figures \ref{fig_1}d, \ref{fig_1}e, and \ref{fig_1}f) with deconvolved major axes, minor axes, and position angles of 1\farcs70 $\pm$ 0\farcs09 ($\sim$ 560~pc), 1\farcs18 $\pm$ 0\farcs13 ($\sim$ 390~pc), and 176\degree.1 $\pm$ 2\degree.3, respectively (Figure \ref{fig_2} and Table \ref{table_gauss}).
The radial distribution of each emission (Figure \ref{fig_3}a) suggests that the low frequency emission ($\leq$ 108~GHz) is systematically more compact relative to the high frequency emission ($\geq$ 233~GHz).
The peak positions coincide with each other as well as the low frequency emission.

Assuming a circular disk geometry for simplicity, we estimate the inclination of the disk from the axial ratio (i.e., the minor axis divided by the major axis) using,
\begin{equation}
\cos{i} = \frac{a}{b}    
\end{equation}
where $i$ is the inclination of the disk, $a$ is the minor axis, and $b$ is the major axis.
The derived $i$ is listed in Table \ref{table_gauss}.
We found that 4.81, 8.36, and 108~GHz disks are nearly face-on (24.86 $\pm$ 0.02\degree on average), while those of 233, 352, and 691~GHz disks are slightly inclined (46.04 $\pm$ 0.16\degree on average) or more elongated toward the north-south direction.

The difference of the apparent morphology between the low frequency and the high frequency emission is caused by the differences of physics which produces the continuum emission.
Assuming the higher frequency emission arises from the cold dust component and the lower frequency emission arises from the thermal bremsstrahlung and/or the non-thermal synchrotron which are associated with star-forming activities (see \S\ref{sed}), the elongated higher frequency emission suggests that cold dust distribution is more extended toward the north-south direction and larger than the star forming regions.
Deeper continuum observations with shorter baselines are necessary in order to trace more diffuse components.

The recovered fluxes at 352 and 691~GHz in the ALMA observations relative to those in the James Clerk Maxwell Telescope (JCMT; \cite{dun00, d&e01}) observations are 18 $\pm$ 4 and 27 $\pm$ 4 \%, respectively.
Large amounts of the 352 and 691~GHz emission are more extended than the MRS scale ($\sim$ 1.5~kpc).

\begin{figure*}
 \begin{center}
	\includegraphics[width=16cm]{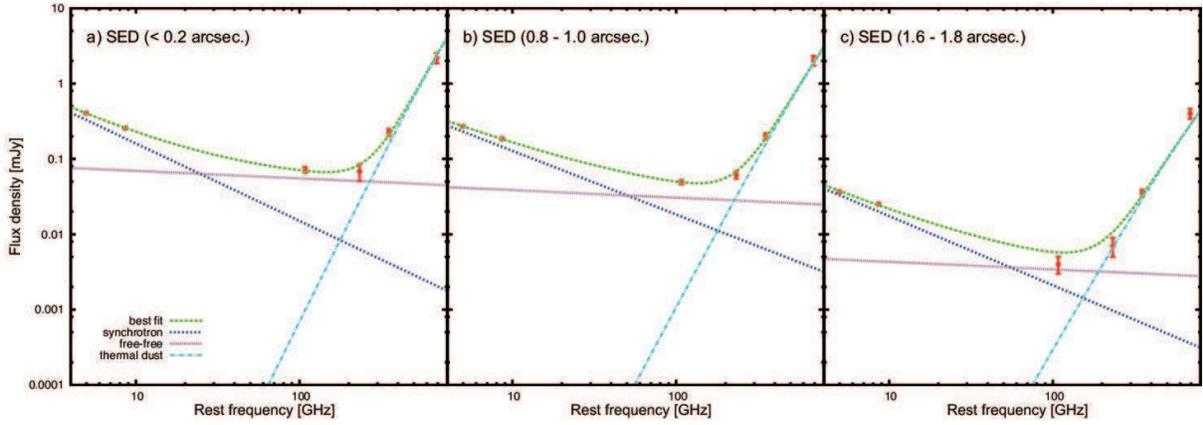}
 \end{center}
\caption{Results of the SED fitting for 3 photometric rings with the radius of (a) 0\farcs0 - 0\farcs2, (b) 0\farcs8 - 1\farcs0, and (c) 1\farcs6 - 1\farcs8.  The blue, purple, and cyan lines show the best-fitted synchrotron, free-free, and dust continuum spectrum, respectively.  The best-fitted spectrum is shown as the green line.  Typical fitting error of the y-axis is 17\%.
}\label{fig_4}
\end{figure*}

\section{SED MODELING} \label{sed}
\subsection{Spatially-resolved radio-to-FIR SED}
We constructed the spatially-resolved SED of NGC~1614 (Figure \ref{fig_3}b) from 4.81 to 691~GHz by measuring the flux densities contained in concentric rings defined by the width of 0\farcs2 (see Figure \ref{fig_3}c).
We note that the flux values between adjacent rings are not independent because the width of the ring is smaller than the convolved beam.
However, since the continuum images are sufficiently resolved with the 1\farcs0 beam, this will not introduce any artificial bias in the discussion of the overall trend presented in the following sections.
The shape of the SEDs are similar from region to region, while the total flux density systematically decreases as the ring radius increases.
The most significant difference is the slope between the 108 and 233 GHz emission (i.e., the nuclear region shows a flatter spectrum).
In order to understand the physical origins of the shape of the SED, we perform a simple SED fitting.

\subsection{SED Formulation for Starburst-dominated Galaxies}
We modeled the radio-to-FIR continuum emission by using the formulation described in \citet{y&c02},
\begin{eqnarray}
S(\nu_{\rm{obs}}) & = & (S_{\rm{nth}} + S_{\rm{ff}} + S_{\rm{d}})(1 + z)\:\:\rm{Jy} \\
S_{\rm{nth}} & = & 25 f_{\rm{nth}}\nu_0^{-\alpha}\frac{\rm{SFR}}{M_{\odot}\:\rm{yr^{-1}}}D_{\rm{L}}^{-2} \\
S_{\rm{ff}} & = & 0.71\nu_0^{-0.1}\frac{\rm{SFR}}{M_{\odot}\:\rm{yr^{-1}}}D_{\rm{L}}^{-2} \label{eq_3}\\
S_{\rm{d}} & = & 1.3 \times 10^{-6} \frac{\nu_0^3\left[1 - e^{-(\nu_0/\nu_c)^{\beta}}\right]}{e^{(0.048\nu_0/T_d)} - 1}\frac{\rm{SFR}}{M_{\odot}\:\rm{yr^{-1}}}D_{\rm{L}}^{-2},
\end{eqnarray}
where $S_{\rm{nth}}$, $S_{\rm{ff}}$, and $S_{\rm{d}}$ are the flux of non-thermal synchrotron, thermal bremsstrahlung (free-free), and thermal dust emission, respectively, $\nu_{\rm{0}}$ is the rest frequency in GHz, $z$ is the redshift, $f_{\rm{nth}}$ is the radio-FIR normalization term adopted by \citet{cnd92}, $\alpha$ is the spectral index of non-thermal synchrotron emission, SFR is the star formation rate in $M_{\odot}$ yr$^{-1}$, $D_{\rm{L}}$ is the luminosity distance, $\nu_c$ is the critical frequency where the clouds become optically thick, $\beta$ is the dust emissivity index, and $T_d$ is the dust temperature.

This formulation is only valid for starburst-dominated galaxies (i.e., AGN contribution to the radio-to-FIR SED is negligible), because it is based on the assumption that all the three components are associated with massive star-forming activities (e.g., supernovae and their remnants, H$\:_{\rm{II}}$ regions surrounding young stars, and dust particles heated by young massive stars).
This assumption is applicable to NGC~1614 since multi-wavelength studies have so far found no significant evidence for an AGN \citep{aln01, i&n13, hi14, xu15}.
In order to minimize the number of free parameters, we fix z = 0.015938, $D_{\rm{L}}$ = 67.8~Mpc \citep{xu15}, $\nu_c$ = 2~THz \citep{y&c02}, and $T_d$ = 35 $\pm$ 2~K \citep{xu15} which is derived by using Spitzer/MIPS 24 $\mu$m and Herschel 70 - 500 $\mu$m data, in the fitting routine.

\begin{figure*}
 \begin{center}
	\includegraphics[width=16cm]{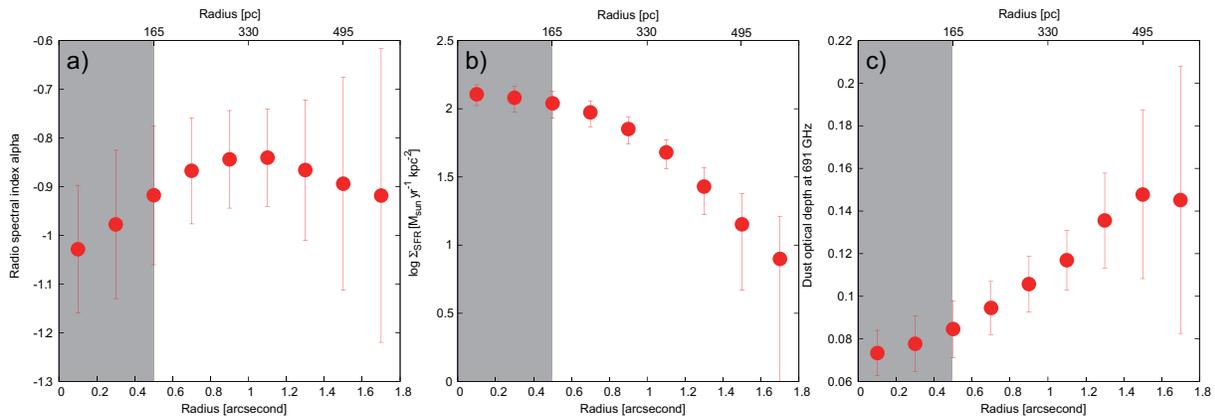}
 \end{center}
\caption{Results of the SED fitting, (a) spectral index $\alpha$, (b) $\Sigma_{\rm{SFR}}$, and (c) dust optical depth at 691~GHz, as a function of radius.  The shaded area indicates the convolved beam size.
}\label{fig_5}
\end{figure*}

The results from the fit are listed in Table \ref{table_sed} and shown in Figure \ref{fig_4}.
The derived $f_{\rm{nth}}$ ranges between 0.43 - 1.45 (averaged $f_{\rm{nth}}$ = 0.66 $\pm$ 0.13).
This is consistent with $f_{\rm{nth}}$ derived by global SED fitting, which typically shows the value of order unity \citep{y&c02}.
The synchrotron spectral index $\alpha$ shows a constant index of $-$0.9 within the error (Figure \ref{fig_5}a).
This is also consistent with global fitting values (e.g., \cite{y&c02}).
The best-fit SFR surface density ($\Sigma_{\rm{SFR}}$) ranges between 10$^{0.9}$ - 10$^{2.1}$ $M_{\odot}$ yr$^{-1}$ kpc$^2$ (Figure \ref{fig_5}b), and this is similarly consistent with the $\Sigma_{\rm{SFR}}$ derived from the extinction-corrected Paschen $\alpha$ measurements \citep{hi14} and the thermal and non-thermal radio measurements decomposed using the 8.36 GHz continuum, near-IR extinction, and Paschen $\alpha$ equivalent width image (see Appendix A of \cite{xu15}), when we take into account the missing flux of the radio-to-FIR SED.
The SFR is mainly determined by $S_{\rm{ff}}$ which dominates the 108~GHz flux, because $S_{\rm{ff}}$ is only proportional to the SFR (Equation \ref{eq_3}).

We calculated the dust optical depth at 691~GHz by $\tau_{\nu}$ = ($\nu$/$\nu_c$)$^{\beta}$ in order to evaluate the contribution of dust extinction to the CO~(6--5) emission.
It is known that some U/LIRGs suffer from strong dust extinction in high-$J$ CO emission (e.g., $\tau_{\rm{350GHz}}$ $>$ 1 for Arp~220; \cite{sak08, ran11}).
In contrast to these studies, we found that the $\tau_{\rm{691GHz}}$ is optically thin (0.06 - 0.21) in the nuclear region of NGC~1614 (central 1\farcs8) suggesting that the dust extinction for the CO~(6--5) emission is negligible (Figure \ref{fig_5}c).

\subsection{Dust Mass and Gas Surface Density Derivation} \label{dust}
We calculate the dust mass from the 352 GHz (852 $\mu$m) continuum emission assuming that the 352~GHz flux is dominated by the cold dust component (Figure \ref{fig_4}) and using the following equation \citep{wil08},
\begin{equation}
M_{\rm{dust}} = 74200\:S_{\rm{352}}\:D_{\rm{L}}^2\:\frac{e^{17/T_d} - 1}{\kappa_{352}}\:\:M_{\rm{\odot}}
\end{equation}
where $S_{\rm{352}}$ is the 352 GHz flux in Jansky, $D_{\rm{L}}$ is the luminosity distance in Mpc, and $\kappa_{352}$ is the dust absorption coefficient (0.43 $\pm$ 0.04 cm$^2$ g$^{-1}$; \cite{l&d01}).
We assumed $T_{\rm{d}}$ of 35 $\pm$ 2 K \citep{xu15}.  The derived dust masses range between 3.3 $\times$ 10$^5$ and 2.6 $\times$ 10$^6$ $M_{\odot}$ (Table \ref{table_sed}).
We note that we used the \citet{l&d01} dust model, because the $\kappa_{352}$ derived from the SED fitting has a large error and the assumption of the constant $T_{\rm{d}}$ of 35 $\pm$ 2 K may affect the derived $\beta$.
High-resolution high-frequency observations are required to constrain the $T_{\rm{d}}$ gradient in the nuclear region of NGC~1614.

Assuming a gas-to-dust ratio of 264 $\pm$ 68 \citep{wil08}, the surface densities of molecular gas ($\Sigma_{\rm{H_2}}$) are estimated to be in the range 10$^{3.0}$ - 10$^{3.8}$ $M_{\odot}$ pc$^{-2}$.
We assumed the CO-to-H$_2$ conversion factor known to be appropriate for U/LIRGs ($\alpha_{\rm{CO}}$ = 0.8 $M_{\odot}$ (K km s$^2$)$^{-1}$; \cite{bol13}).
This is consistent with the value derived by \citet{slw14} in NGC~1614 ($\alpha_{\rm{CO}}$ = 0.9 - 1.5 $M_{\odot}$ (K km s$^2$)$^{-1}$).
The derived $\Sigma_{\rm{H_2}}$ are consistent with the values estimated from the high-resolution CO~(6--5) image \citep{xu15}.
Here, we ignore the $S_{\rm{nth}}$ and the $S_{\rm{ff}}$ contributions to the 352~GHz flux (although \cite{wil08} assumed these contribution by a factor of 0.6 to derive the gas-to-dust ratio of 440 $\pm$ 114), since our SED fittings yield that the 352~GHz flux is dominated by the cold dust emission (Figure \ref{fig_4}).

\section{MERGER EVOLUTION ON THE STAR-FORMING RELATION}
\begin{figure}
 \begin{center}
	\includegraphics[width=8cm]{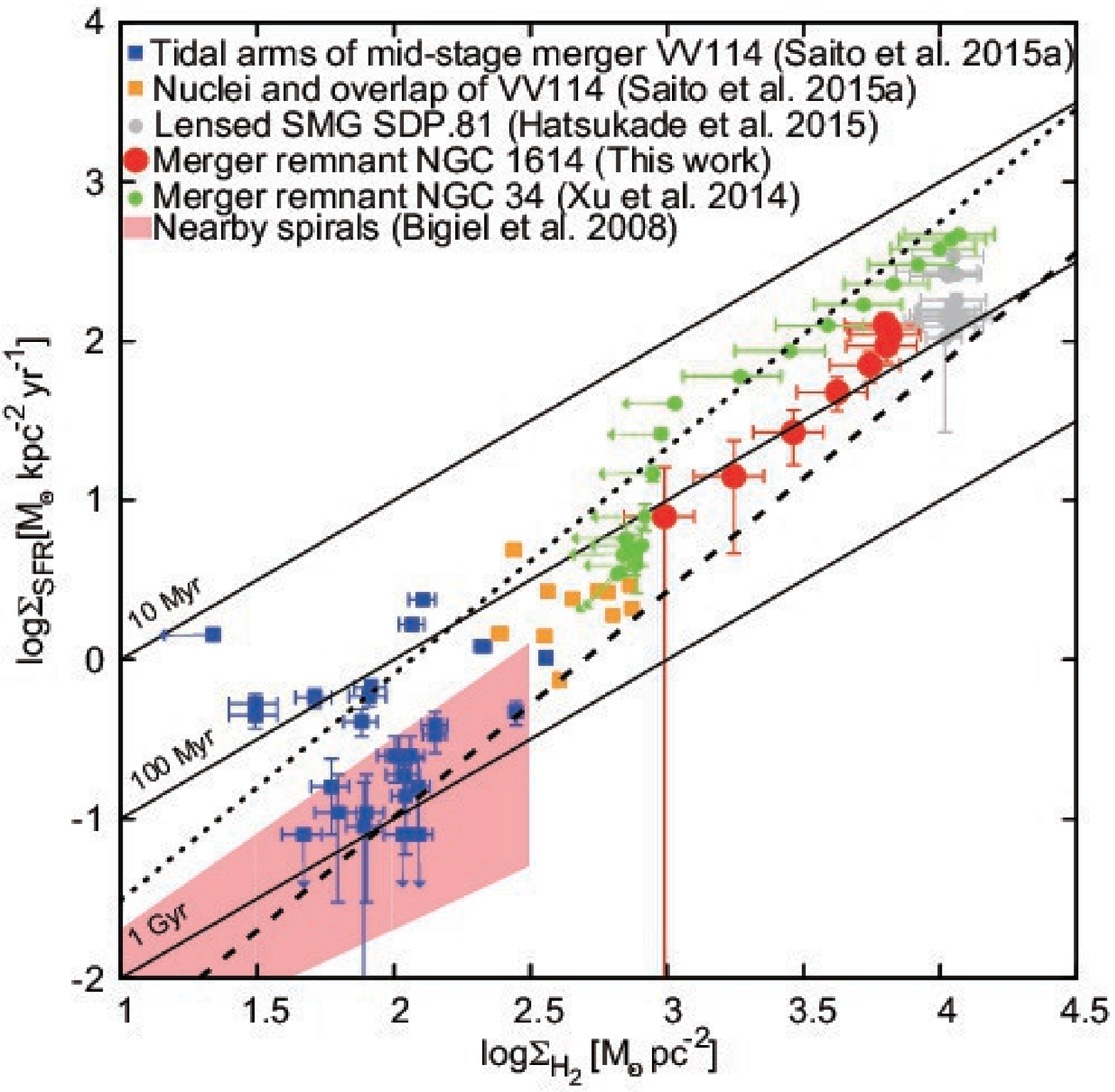}
 \end{center}
\caption{The spatially-resolved (sub-kpc) Kennicutt-Schmidt relation of merger-remnants NGC~1614 (red; this work), NGC~34 (green; \cite{xu14}), a mid-stage merger VV~114 (blue and yellow; \cite{TS15}), a lensed SMG SDP.81 (grey; \cite{hat15}), and nearby spirals (pink shaded area; \cite{big08}).
The solid lines show a constant gas depletion time ($\tau_{\rm{gas}}$) of 10~Myr, 100~Myr, and 1~Gyr.
The dotted and dashed line indicate the ``starburst" sequence and the ``normal disk" sequence \citep{dad10}.
}\label{fig_6}
\end{figure}

We present the sub-kpc scale star-forming (Kennicutt-Schmidt; KS) relation for NGC~1614 (Figure \ref{fig_6}), and compare with similar spatial resolution (0.3 - 0.8 kpc) data of U/LIRGs in the literature (VV~114, NGC~34, and SDP.81; \cite{xu14, hat15, TS15}).
We also make a comparison with $\sim$ 0.75~kpc resolution data of nearby spiral galaxies \citep{big08}.

NGC~1614 (red points in Figure \ref{fig_6}) has higher $\Sigma_{\rm{H_2}}$ (= 10$^{3.0 - 3.8}$ $M_{\odot}$ pc$^{-2}$) and $\Sigma_{\rm{SFR}}$ (= 10$^{0.9 - 2.1}$ $M_{\odot}$ yr$^{-1}$ kpc$^{-2}$) than nearby spirals.
In addition, we derive the molecular gas depletion time ($\tau_{\rm{gas}}$ = $\Sigma_{\rm{H_2}}$/$\Sigma_{\rm{SFR}}$ yr) which is the time needed to consume the existing molecular gas by star formation.
The nuclear region of NGC~1614 ($<$ 300~pc) has shorter gas depletion time (averaged $\tau_{\rm{gas}}$ = 49 - 77~Myr) than the outer region (averaged $\tau_{\rm{gas}}$ = 74 - 226~Myr) (Table~\ref{table_sed}).
The region with the shorter $\tau_{\rm{gas}}$ coincides with the starburst ring \citep{aln01}, which suggests that most of the intense SB in NGC~1614 occur within the ring rather than the outer region.

VV~114 is a gas-rich mid-stage merging galaxy with an infrared luminosity of 10$^{11.71}$ $L_{\odot}$ \citep{arm09}.
Although the system has a possible buried AGN as identified by dense molecular gas tracers \citep{ion13} and the unresolved hard X-ray emission \citep{grm06}, the extended infrared continuum, Paschen $\alpha$ \citep{tat15}, and low-$J$ CO emission \citep{ion04} mainly come from kpc-scale clumpy star-forming regions.
We used the ALMA CO~(1--0) and the miniTAO Paschen $\alpha$ emission to derive the $\Sigma_{\rm{H_2}}$ and $\Sigma_{\rm{SFR}}$ of VV~114, respectively (see \cite{TS15} for details).
VV~114 has the lowest star formation among the four galaxies shown in Figure \ref{fig_6}.
The blue points indicate star-forming regions in tidal arms, and the yellow points indicate the nuclei and overlap region between the galaxy disks.
Most of SF regions have the $\tau_{\rm{gas}}$ between 100~Myr and 1~Gyr.
These values are shorter than normal spiral galaxies \citep{big08}, and they are slightly higher than the ``normal disk" sequence (index = 1.4) which is mainly dominated by nearby spiral galaxies with moderate star formation \citep{dad10}.

NGC~34 (green points in Figure \ref{fig_6}) is one of the bright merger remnants ($L_{\rm{IR}}$ = 10$^{11.49}$ $L_{\odot}$; \cite{arm09}) with a rotating cold gas disk ($R$ $\sim$ 3~kpc; \cite{ued14}) and warm compact gas disk ($R$ $\sim$ 320 pc; \cite{xu14}) similar to NGC~1614.
We used the ALMA 435~$\mu$m dust and the VLA 8.44~GHz emission to derive the $\Sigma_{\rm{H_2}}$ and $\Sigma_{\rm{SFR}}$ of NGC~34, respectively (see \cite{xu14} in details).
NGC~34 has a systematically higher $\Sigma_{\rm{SFR}}$ than NGC~1614 (i.e., $\tau_{\rm{gas}}$ of NGC~34 is shorter than that of NGC~1614).
As suggested by \citet{xu15}, one possibility to explain the difference is the characteristics of the progenitor galaxies.
NGC~1614 is known as a minor merger, a colliding system between a massive and a dwarf galaxy with the mass ratio of $>$ 4 \citep{vai12}, whereas NGC~34 is known as a major merger \citep{xu14}.
Since major mergers efficiently induce gas inflows toward their nuclei due to the strong tidal interaction (e.g., \cite{cox08}), the central starbursts in major mergers can be more intense than that of minor mergers.

SDP.81 (grey points in Figure \ref{fig_6}) is one of the brightest lensed submillimeter galaxies (z = 3.042; Intrinsic $L_{\rm{FIR}}$ = 10$^{12.70}$ $L_{\odot}$) discovered in the Herschel Astrophysical Terahertz Large Area Survey \citep{eal10}.
Using a gravitational lens model, \citet{hat15} found evidence of a rotating molecular gas disk with spatially-decoupled stellar components which suggests central dusty starbursts or merging signatures between two gas-rich galaxies (i.e., tidal tails or arms).
\citet{hat15} used the ALMA CO~(5--4) and the ALMA rest-frame 125 $\mu m$ emission to derive the $\Sigma_{\rm{H_2}}$ and $\Sigma_{\rm{SFR}}$ of SDP.81, respectively.
The $\tau_{\rm{gas}}$ of SDP.81 ($\sim$ 50~Myr) is longer than that of NGC~34 and similar to that of the NGC~1614 nucleus.
All of the clumps that are distributed over the kpc-scale disk of SDP.81 show active SF, which is similar to the nuclear region of NGC~1614 and NGC~34 (central 200~pc).

Overall, the sub-kpc KS relation for U/LIRGs has a larger index than unity ($\sim$ 1.4; Figure \ref{fig_6}).
The index is larger than the fitted value for spatially-resolved (sub-kpc) nearby spiral galaxies (1.0 $\pm$ 0.2; \cite{big08}), which suggests that molecular clumps in U/LIRGs (especially their nuclei) form stars with higher efficiency than in spirals.
Furthermore, although the sample size is not enough statistically, our U/LIRG sample is distributed between the ``normal disk" and the ``starburst" sequence both of which have a same index of 1.4 \citep{dad10}.
This is consistent with a numerical simulation provided by \citet{bou11} who suggest that galaxy mergers between two gas-rich disks can evolve toward  the ``starburst" mode (i.e., the $\tau_{\rm{gas}}$ becomes shorter as the merger-stage proceeds from VV~114 to NGC~1614 to NGC~34).

Finally we note that, since the $\Sigma_{\rm{H_2}}$ of all the U/LIRGs are derived by simply assuming a CO luminosity-to-H$_2$ mass conversion factor ($\alpha_{\rm{CO}}$) of 0.8 $M_{\odot}$ (K km s$^{-1}$ pc$^2$)$^{-1}$, which is the standard value for U/LIRGs (e.g., \cite{bol13}), and/or adopting observed gas-to-dust ratio, data points in Figure \ref{fig_6} systematically shift when we adopt other $\alpha_{\rm{CO}}$.
Therefore, the index of 1.4 does not change significantly.

\section{CONCLUSION}
Our new cycle 2 ALMA observations toward NGC~1614 of the 108 and 233~GHz continuum, combined with archival cycle 0 ALMA data of the 352 and 691~GHz continuum, and archival VLA data of the 4.81 and 8.36~GHz emission show a similar structure between each emission, although there are systematic differences between the lower frequency emission ($\leq$ 108~GHz) and the higher frequency emission ($\geq$ 233~GHz).
The higher frequency emission has a more extended distribution toward the north-south direction, while both the lower and higher frequency emission show an offset peak relative to the nucleus, which was detected in the higher-resolution 5~GHz and Paschen $\alpha$ images, and a ring-like residual after subtracting the fitted 2D gaussian.
This is the first result to resolve the radio-to-FIR continuum emission in LIRGs with 1\farcs0 resolution simultaneously.

Using the starburst SED template, we confirmed that the spatial differences of the SED may be due to the differences of dominant emitting mechanism; i.e., (1) 4.81 and 8.36~GHz fluxes are dominated by the non-thermal synchrotron, (2) 108~GHz flux is dominated by the thermal free-free emission, (3) 352 and 691~GHz fluxes are dominated by the thermal dust emission.

We also estimated the spatially-resolved (1\farcs0 $\times$ 1\farcs0) star formation rate surface densities and molecular gas surface densities.
By combining with data in the literature, we suggest a trend that the molecular gas depletion time decreases as galaxy merger stage proceeds.
This is consistent with the prediction from numerical simulations of star-forming activities as a function of the merger stage.

In forthcoming papers, we will provide underlying gas physics (e.g., excitation) and chemistry (e.g., abundance) in merger-induced starbursts with high sensitivity and resolution (0\farcs2 - 1\farcs0) enough to resolve the starburst ring in NGC~1614 (Ando et al. in prep. and Saito et al. in prep.).

\begin{table*}
  \tbl{Imaging properties of continuum emission.}{%
  \begin{tabular}{ccccccccc}
      \hline
      Telescope &Band &$\nu_{\rm{rest}}$ &MRS$^*$ &$uv$-weight$\dagger$ &Beam size$\ddagger$ &$S_{\rm{\nu}}$$\S$ &Recovered flux$\|$ &Ref.$\#$ \\
      & &[GHz] &[\arcsec] & &[\arcsec] &[mJy] &[\%] & \\
      \hline
      VLA &C &4.81 &4.6 &uniform &1.0 $\times$ 1.0 &26.69 $\pm$ 0.13 &42 $\pm$ 8 &1, a \\
      VLA &X &8.36 &4.6 &briggs &1.0 $\times$ 1.0 &17.66 $\pm$ 0.07 &... &b \\
      ALMA &B3 &108 &4.6 &briggs &1.0 $\times$ 1.0 &\phantom{0}6.39 $\pm$ 0.13 &... &This work \\
      ALMA &B6 &233 &4.6 &uniform &1.0 $\times$ 1.0 &\phantom{0}8.02 $\pm$ 0.36 &... &This work \\
      ALMA &B7 &352 &4.6 &briggs &1.0 $\times$ 1.0 &29.23 $\pm$ 0.23 &18 $\pm$ 4 &2, c \\
      ALMA &B9 &691 &4.6 &briggs &1.0 $\times$ 1.0 &222.3 $\pm$ 11.2 &27 $\pm$ 4 &3, d \\
      \hline
    \end{tabular}}\label{table_contin}
\begin{tabnote}
      \par\noindent
      \footnotemark[$*$] Maximum recoverable scale (MRS) of each observations.  This is defined by $\sim$ 0.6 $\lambda_{\rm{obs}}$/(minimum baseline length), where $\lambda_{\rm{obs}}$ is the observed wavelength.
      \par\noindent
      \footnotemark[$\dagger$] Visibility ($uv$-plane) weighting for imaging.  ``briggs" means Briggs weighting with {\tt robust} = 0.5.
      \par\noindent
      \footnotemark[$\ddagger$] The synthesized beam size.  The images are convolved into the same resolution.
      \par\noindent
      \footnotemark[$\S$] Integrated flux density enclosed with the 3$\sigma$ contour.  We only consider the statistical error in this column.  The systematic error of absolute flux calibration is estimated to be 3\%, 3\%, 5\%, 10\%, 10\%, and 15\% for VLA/C. VLA/X, ALMA/B3, ALMA/B6, ALMA/B7, and ALMA/B9, respectively.
      \par\noindent
      \footnotemark[$\|$] The ALMA/VLA flux divided by the single-dish flux.  We consider the statistical and systematic error.
      \par\noindent
      \footnotemark[$\#$] Reference of the single-dish flux (1 = \citet{bic95}, 2 = \citet{dun00}, and 3 = \citet{d&e01}) and the interferometric data (a = \citet{ols10}, b = \citet{sch06}, c = \citet{slw14}, and d = \citet{xu15}).
\end{tabnote}
\end{table*}

\begin{table*}
  \tbl{Results of the 2D gaussian fitting.}{%
  \begin{tabular}{ccccccc}
      \hline
      Band &R.A. &Decl. &$R_{\rm{deconv}}$$*$ &P.A. &Axis ratio$\dagger$ &$i$$\ddagger$ \\
      & [h m s] &[d m s] &[\arcsec] &[\degree] & &[\degree] \\
      \hline
      C &4:34:00.013 &$-$8.34.44.918 &1.279 $\pm$ 0.006 &\phantom{0}60.56 $\pm$ 0.49 &0.918 $\pm$ 0.006 &23.41 $\pm$ 0.04 \\
      X &4:34:00.012 &$-$8.34.45.057 &1.214 $\pm$ 0.006 &115.49 $\pm$ 0.40 &0.926 $\pm$ 0.007 &22.16 $\pm$ 0.05 \\
      B3 &4:34:00.017 &$-$8.34.45.096 &1.350 $\pm$ 0.026 &176.2 $\pm$ 2.1 &0.859 $\pm$ 0.027 &30.85 $\pm$ 0.05 \\
      B6 &4:34:00.013 &$-$8.34.45.030 &1.626 $\pm$ 0.058 &\phantom{00}1.4 $\pm$ 2.1 &0.734 $\pm$ 0.048 &42.75 $\pm$ 0.07 \\
      B7 &4:34:00.014 &$-$8.34.44.959 &1.781 $\pm$ 0.015 &171.88 $\pm$ 0.48 &0.728 $\pm$ 0.011 &43.26 $\pm$ 0.02 \\
      B9 &4:34:00.006 &$-$8.34.45.103 &1.681 $\pm$ 0.063 &174.89 $\pm$ 0.78 &0.630 $\pm$ 0.050 &50.95 $\pm$ 0.08 \\
      \hline
    \end{tabular}}\label{table_gauss}
\begin{tabnote}
      \par\noindent
      \footnotemark[$*$] Deconvolved source size which is defined as [$R_{\rm{conv}}$ - ($\theta_{\rm{HPBW}}$/2)$^2$]$^{0.5}$, where $R_{\rm{conv}}$ is the estimated major axis size of the gaussian fitting and $\theta_{\rm{HPBW}}$ is the synthesized beam size.
      \par\noindent
      \footnotemark[$\dagger$] Value which the minor axis divided by the major axis.
      \par\noindent
      \footnotemark[$\ddagger$] Inclination with an assumption of a thin disk. See text.
\end{tabnote}
\end{table*}

\begin{table*}
  \tbl{Results of the SED fitting.}{%
  \begin{tabular}{cccccccc}
      \hline
      Radius$*$ &$f_{\rm{nth}}$$\dagger$ &$\alpha$$\ddagger$ &$\beta$$\S$ &log $\Sigma_{\rm{SFR}}$ &$M_{\rm{dust}}$ &log $\Sigma_{\rm{H_2}}$$\|$ &log $\tau_{\rm{gas}}$$\#$ \\
      & & & &[$M_{\odot}$ yr$^{-1}$ kpc$^{-2}$] &[$\times$ 10$^5$ $M_{\odot}$] &[$M_{\odot}$ pc$^{-2}$] &[yr] \\
      \hline
      0\farcs0 - 0\farcs2 &0.56 $\pm$ 0.09 &$-$1.03 $\pm$ 0.13 &2.46 $\pm$ 0.16 &2.11$^{+0.07}_{-0.08}$ &\phantom{0}3.3 $\pm$ 0.4 &3.80$^{+0.11}_{-0.15}$ &7.69$^{+0.19}_{-0.22}$ \\
      0\farcs2 - 0\farcs4 &0.54 $\pm$ 0.09 &$-$0.98 $\pm$ 0.15 &2.40 $\pm$ 0.19 &2.08$^{+0.08}_{-0.10}$ &10.1 $\pm$ 1.3 &3.81$^{+0.11}_{-0.15}$ &7.73$^{+0.21}_{-0.23}$ \\
      0\farcs4 - 0\farcs6 &0.51 $\pm$ 0.07 &$-$0.92 $\pm$ 0.14 &2.32 $\pm$ 0.19 &2.04$^{+0.09}_{-0.11}$ &17.1 $\pm$ 2.2 &3.82$^{+0.11}_{-0.15}$ &7.78$^{+0.22}_{-0.24}$ \\
      0\farcs6 - 0\farcs8 &0.50 $\pm$ 0.07 &$-$0.87 $\pm$ 0.11 &2.22 $\pm$ 0.17 &1.97$^{+0.08}_{-0.10}$ &23.2 $\pm$ 2.9 &3.81$^{+0.11}_{-0.15}$ &7.84$^{+0.21}_{-0.23}$ \\
      0\farcs8 - 1\farcs0 &0.52 $\pm$ 0.07 &$-$0.84 $\pm$ 0.10 &2.11 $\pm$ 0.16 &1.85$^{+0.09}_{-0.11}$ &25.9 $\pm$ 3.3 &3.74$^{+0.11}_{-0.15}$ &7.89$^{+0.22}_{-0.24}$ \\
      1\farcs0 - 1\farcs2 &0.56 $\pm$ 0.08 &$-$0.84 $\pm$ 0.10 &2.02 $\pm$ 0.17 &1.68$^{+0.09}_{-0.12}$ &24.0 $\pm$ 3.0 &3.62$^{+0.11}_{-0.15}$ &7.94$^{+0.23}_{-0.24}$ \\
      1\farcs2 - 1\farcs4 &0.70 $\pm$ 0.16 &$-$0.87 $\pm$ 0.14 &1.88 $\pm$ 0.26 &1.43$^{+0.14}_{-0.21}$ &19.6 $\pm$ 2.5 &3.46$^{+0.11}_{-0.15}$ &8.03$^{+0.32}_{-0.29}$ \\
      1\farcs4 - 1\farcs6 &0.80 $\pm$ 0.40 &$-$0.89 $\pm$ 0.22 &1.80 $\pm$ 0.46 &1.15$^{+0.22}_{-0.48}$ &13.7 $\pm$ 1.7 &3.25$^{+0.11}_{-0.15}$ &8.11$^{+0.59}_{-0.37}$ \\
      1\farcs6 - 1\farcs8 &0.76 $\pm$ 0.69 &$-$0.92 $\pm$ 0.30 &1.82 $\pm$ 0.72 &0.90$^{+0.31}_{-0.90}$ &\phantom{0}8.7 $\pm$ 1.1 &2.99$^{+0.11}_{-0.15}$ &8.09$^{+1.01}_{-0.46}$ \\
      \hline
    \end{tabular}}\label{table_sed}
\begin{tabnote}
      \par\noindent
      \footnotemark[$*$] Inside and outside radii of the concentric ring which are used for the photometry.
      \par\noindent
      \footnotemark[$\dagger$] Multiplicative correction factor for non-thermal radio continuum emission.  The $f_{\rm{nth}}$ = 1.0 corresponds to the Galactic normalization adopted by \citet{cnd92}.
      \par\noindent
      \footnotemark[$\ddagger$] The synchrotron spectral index.
      \par\noindent
      \footnotemark[$\S$] Dust emissivity.
      \par\noindent
      \footnotemark[$\|$] Molecular gas surface density assuming the gas-to-dust ratio of 264 $\pm$ 68. See text.
      \par\noindent
      \footnotemark[$\#$] Gas depletion time which is defined by $\Sigma_{\rm{SFR}}$/$\Sigma_{\rm{H_2}}$ yr.
\end{tabnote}
\end{table*}

\bigskip

The authors thanks the anonymous referee for comments that improved this paper.
T.S. and other authors thank ALMA staff for their kind support.
T.S. and M. Lee are financially supported by a Research Fellowship from the Japan Society for the Promotion of Science for Young Scientists.
T.S. was supported by the ALMA Japan Research Grant of NAOJ Chile Observatory, NAOJ-ALMA-0114.
D. Iono was supported by the ALMA Japan Research Grant of NAOJ Chile Observatory, NAOJ-ALMA-0011, JSPS KAKENHI Grant Number 15H02074, and the 2015 Inamori Research Grants Program.
This paper makes use of the following ALMA data: ADS/JAO.ALMA\#2011.0.00182.S, ADS/JAO.ALMA\#2011.0.00768.S, and ADS/JAO.ALMA\#2013.1.01172.S.
ALMA is a partnership of ESO (representing its member states), NSF (USA) and NINS (Japan), together with NRC (Canada), NSC and ASIAA (Taiwan), and KASI (Republic of Korea), in cooperation with the Republic of Chile.
The joint ALMA Observatory is operated by ESO, AUI/NRAO, and NAOJ.
This research has made extensive use of the NASA/IPAC Extragalactic Database (NED) which is operated by the Jet Propulsion Laboratory, California Institute of Technology, under contract with the National Aeronautics and Space Administration.


\end{document}